\documentclass[%
reprint,
onecolumn,
 amsmath,amssymb,
 aps,
]{revtex4-2}

\usepackage{graphicx}
\usepackage{bm}
\usepackage{braket}
\usepackage{amsmath}

\begin{document}

\title{Separation out of Entanglement}


\author{Toru Ohira}
 \altaffiliation
{Graduate School of Mathematics, Nagoya University, Nagoya, Japan
}%

\date{\today}

\begin{abstract}%
We investigate the separability properties of quantum states described by an extended Werner density matrix, where the classical component exhibits statistical dependence. By generalizing the classical part to allow correlations, we demonstrate that within a specific parameter range the separable region expands compared to the standard Werner state with an independent classical component. This result suggests that increasing classical correlation can enhance the separability of the overall quantum state, providing new insights into the interplay between classical and quantum correlations.
\end{abstract}

\maketitle

\section{Introduction}
We often encounter entangled situations in everyday life, such as tangled computer and audio cables,  
human relationships ranging from personal to international politics, interest rates and currency exchange, and so on.  
In the realm of quantum mechanics and its applications, entanglement has also become a fundamental and widely studied concept.  

Hence, the determination of whether a quantum system is separable or entangled has been a subject of active research  
(e.g., \cite{bell,clauser,aspect,leggett,vedral,guhne03,altepeter,sakai,werner,peres,horodecki}).  
This question becomes particularly challenging when the quantum state is mixed rather than pure,  
which remains a central focus of ongoing efforts to establish conditions for the separability of density matrices  
(e.g., \cite{werner,peres,horodecki,horodecki2}).  

Now, imagine placing two sets of entangled strings into a box and mixing them together - what would we get?  
In classical systems, we would likely end up with an even more entangled mess.  
However, in the quantum world, the outcome can be quite different.  
What we demonstrate in this letter is that when two components, each individually correlated and entangled, are combined in a suitable manner,  
a separable quantum state can emerge.  

We illustrate this phenomenon using a quantum state described by the extended Werner density matrix.  
Although the Werner density matrix represents the simplest example of a two-particle (bipartite)  
$2\times 2$ system in a mixed state, it provides valuable insights into various aspects of quantum mechanics.  
Quantum states characterized by the Werner density matrix have been extensively studied, including investigations of concurrence measures \cite{chen}, quantum tomography, nonlocality \cite{vertesi}, entanglement purification \cite{bennett},  
and quantum teleportation \cite{lee,yao}. These states can also be experimentally realized using polarization-entangled photon pairs \cite{barbieri}.  

The standard Werner matrix consists of an independent and unbiased classical component and a quantum entangled singlet state.  
These components are linearly combined using a controlling parameter, $\xi$.  
By applying the Peres-Horodecki separability criterion, we find that there exists a critical value of $\xi_c = 1/3$.  
When the independent classical component is mixed above $1 - \xi_c = 2/3$, the state remains separable.  
However, if the quantum component dominates beyond this threshold $\xi_c$, the overall state becomes entangled.  

Here, we extend the Werner matrix to include a classically dependent component, such that the classical part exhibits correlations while the quantum part retains entanglement.
Interestingly, we find that an appropriate mixture can still result in a separable state.
Moreover, the critical control value increases to $\xi_c = 1/2$, indicating an enhanced degree of separability.  
Thus, we demonstrate that a separable mixed state can be constructed from components that are classically correlated and quantum-mechanically entangled.


\section{Werner Density Matrix}

The Werner density matrix describes quantum bipartite systems of two 2-state particles ($2 \times 2$ system), such as two spin-1/2 particles or two qubits (e.g., \cite{werner,wootters,kummer,abouraddy,jchen,fujikawa2}).

It is known that any density matrix for $2 \times 2$ systems can be expressed using Pauli matrices as follows:

\begin{equation}
\rho_{AB} = \frac{1}{4}(\mathbf{1}_A \otimes \mathbf{1}_B + \vec{a} \cdot \vec{\mathcal{\sigma}} \otimes \mathbf{1}_B + \mathbf{1}_A \otimes \vec{b} \cdot \vec{\mathcal{\sigma}} + \sum_{ij} F_{ij}\mathcal{\sigma}_i \otimes \mathcal{\sigma}_j)
\label{density}
\end{equation}
where $\mathbf{1}$ is the $2 \times 2$ identity matrix, $\vec{a}$ and $\vec{b}$ are vectors consisting of three real numbers (with $\cdot$ denoting the inner product), and $F_{ij}$ are the real elements of a $3 \times 3$ matrix $\mathcal{F}$. The vector $\vec{\mathcal{\sigma}} = (\mathcal{\sigma}_x, \mathcal{\sigma}_y, \mathcal{\sigma}_z)$ represents the Pauli matrices:

\begin{equation}
\mathcal{\sigma}_x =
\begin{bmatrix}
0 & 1 \\
1 & 0
\end{bmatrix}, \quad
\mathcal{\sigma}_y =
\begin{bmatrix}
0 & -i \\
i & 0
\end{bmatrix}, \quad
\mathcal{\sigma}_z =
\begin{bmatrix}
1 & 0 \\
0 & -1
\end{bmatrix}.
\end{equation}

The Werner density matrix is a mixture of the density matrix describing the $2 \times 2$ quantum singlet state and the density matrix for a classical, independent, and unbiased $2 \times 2$ system.

First, the $2 \times 2$ singlet state, which is one of the Bell states, is given by:

\begin{equation}
\ket{\Psi^{-}} = \frac{1}{\sqrt{2}}(\ket{a_1} \otimes \ket{b_2} - \ket{a_2} \otimes \ket{b_1})
\equiv \frac{1}{\sqrt{2}} \left( \begin{bmatrix} 1 \\ 0 \end{bmatrix}_A \otimes \begin{bmatrix} 0 \\ 1 \end{bmatrix}_B - \begin{bmatrix} 0 \\ 1 \end{bmatrix}_A \otimes \begin{bmatrix} 1 \\ 0 \end{bmatrix}_B \right).
\label{bell2}
\end{equation}

The corresponding density matrix in the notation above is:

\begin{equation}
\rho^{\text{singlet}}_{AB} = \frac{1}{4} \left( \mathbf{1}_A \otimes \mathbf{1}_B + \sum_{i} (-1) \mathcal{\sigma}_i \otimes \mathcal{\sigma}_i \right) = \frac{1}{2} \begin{bmatrix}
0 & 0 & 0 & 0 \\
0 & 1 & -1 & 0 \\
0 & -1 & 1 & 0 \\
0 & 0 & 0 & 0
\end{bmatrix}.
\label{density-singlet}
\end{equation}

On the other hand, a classical, unbiased, independent $2 \times 2$ system behaves like two independent ideal coin flips, corresponding to a classical bipartite state where both $A$ and $B$ have two states, each with equal probabilities of $1/2$. The density matrix for such a system is simply:
\begin{equation}
\rho^{\text{unbiased}}_{AB} = \frac{1}{4} \mathbf{1}_A \otimes \mathbf{1}_B = \frac{1}{4} \begin{bmatrix}
1 & 0 & 0 & 0 \\
0 & 1 & 0 & 0 \\
0 & 0 & 1 & 0 \\
0 & 0 & 0 & 1
\end{bmatrix}.
\end{equation}

The Werner density matrix for bipartite $2 \times 2$ quantum systems is given by:
\begin{equation}
\rho^{W}_{AB} = \frac{1-\xi}{4} \mathbf{1}_A \otimes \mathbf{1}_B + \xi \ket{\Psi^{-}} \bra{\Psi^{-}} = \frac{1}{4} \begin{bmatrix}
1-\xi & 0 & 0 & 0 \\
0 & 1+\xi & -2\xi & 0 \\
0 & -2\xi & 1+\xi & 0 \\
0 & 0 & 0 & 1-\xi
\end{bmatrix},
\end{equation}
where $0 \leq \xi \leq 1$, and $\ket{\Psi^{-}}$ is the singlet state from equation (\ref{bell2}). When $\xi = 1$, it reduces to the pure singlet state, while for $\xi = 0$, it describes a classical unbiased, independent $2 \times 2$ system. Additionally, it can be written as:

\begin{equation}
\rho^{W}_{AB} = \frac{1}{4} \left( \mathbf{1}_A \otimes \mathbf{1}_B + (-\xi) \sum_{i} \mathcal{\sigma}_i \otimes \mathcal{\sigma}_i \right).
\label{mixdensity}
\end{equation}

Thus, the Werner density matrix can be viewed as a combination of a classical unbiased independent bipartite state and an entangled quantum singlet state, with the parameter $\xi$ controlling the degree of this mixture.

This density matrix has been extensively studied (e.g., \cite{peres,hiroshima,azuma}) and is known to be separable for $0 \leq \xi \leq \frac{1}{3}$, and entangled for $\frac{1}{3} < \xi \leq 1$.

\section{Extended Werner Density Matrix}

We extend the Werner density matrix by allowing the classical component to exhibit dependence. Specifically, we consider two correlated classical two-state systems, characterized by the joint probabilities $P(a, b)$ for the event that $A$ takes state $a$ and $B$ takes state $b$. We define the parameters:

\begin{equation}
P(a_1, b_1) = \alpha, \quad P(a_1, b_2) = \beta, \quad P(a_2, b_1) = \gamma, \quad P(a_2, b_2) = \delta,
\end{equation}
with the constraint $\alpha + \beta + \gamma + \delta = 1$.

This system is described by the following density matrix, which is generally not independent:

\begin{equation}
\rho^{\text{general}}_{AB} = 
\begin{bmatrix}
\alpha & 0 & 0 & 0 \\
0 & \beta & 0 & 0 \\
0 & 0 & \gamma & 0 \\
0 & 0 & 0 & \delta
\end{bmatrix}.
\end{equation}
We note in passing that there is no entanglement present, and the density matrix is itself separable.
However, we emphasize that it represents the most general form of a classical mixed state that is not, in general, statistically independent (see below and Appendix for details on the independent case).

Replacing the classical component in the Werner density matrix with this generalized form, we obtain:

\begin{equation}
\hat{\rho}^{W}_{AB} = (1 - \xi) \rho^{\text{general}}_{AB} 
+ \xi \ket{\Psi^{-}} \bra{\Psi^{-}},
\end{equation}

or, explicitly in matrix form:

\begin{equation}
\hat{\rho}^{W}_{AB} = 
\begin{bmatrix}
\alpha (1-\xi) & 0 & 0 & 0 \\
0 & \beta(1 - \xi) + \frac{\xi}{2} & -\frac{\xi}{2} & 0 \\
0 & -\frac{\xi}{2} & \gamma(1 - \xi) + \frac{\xi}{2} & 0 \\
0 & 0 & 0 & \delta(1 - \xi)
\end{bmatrix}.
\label{eWerner}
\end{equation}

We recover the original Werner density matrix when \( \alpha = \beta = \gamma = \delta = \frac{1}{4} \). Additionally, we note that when  
\begin{equation}
\alpha = p_A p_B, \quad \beta = p_A (1 - p_B), \quad \gamma = (1 - p_A) p_B, \quad \delta = (1 - p_A) (1 - p_B),
\end{equation}
where  

- A takes state \( a_1 \) with probability \( p_A \) and state \( a_2 \) with probability \( 1 - p_A \),  

- B takes state \( b_1 \) with probability \( p_B \) and state \( b_2 \) with probability \( 1 - p_B \),  
 
 \noindent
the classical component remains independent but is generally biased. Furthermore, for \( p_A = p_B = \frac{1}{2} \), we obtain the unbiased independent case, recovering the original Werner density matrix. (See the Appendix for details.)

\section{Separability out of Non-inependence and Entanglement}

We examine how this generalization of the Werner density matrix affects its entanglement properties. The Peres-Horodecki criterion states that a bipartite density matrix is separable if and only if all the eigenvalues of its partially transposed matrix are non-negative; otherwise, the system is entangled. We proceed by applying this criterion.

The partially transposed matrix corresponding to this extended matrix is given as follows:

\begin{equation}
\hat{{\rho}}^{W^T}_{AB} =
\begin{bmatrix}
\alpha(1 - \xi) & 0 & 0 & -\frac{\xi}{2} \\
0 & \beta(1 - \xi) + \frac{\xi}{2} & 0 & 0 \\
0 & 0 & \gamma(1 - \xi) + \frac{\xi}{2} & 0 \\
-\frac{\xi}{2} & 0 & 0 & \delta(1 - \xi)
\end{bmatrix}.
\label{eWernerTrans}
\end{equation}

The characteristic polynomial of this matrix is given by:
\begin{equation}
g(\lambda) = \left( \beta(1 - \xi) + \frac{\xi}{2} - \lambda \right) 
\left( \gamma(1 - \xi) + \frac{\xi}{2} - \lambda \right) 
\left[ \lambda^2 - (\alpha + \delta)(1 - \xi) \lambda + \left( \alpha \delta(1 - \xi)^2 - \frac{\xi^2}{4} \right) \right].
\end{equation}
The eigenvalues are the solutions of $g(\lambda) = 0$. Given the conditions $0 \leq \alpha, \beta, \gamma, \delta \leq 1$ and $0 \leq \xi \leq 1$, we obtain the following eigenvalues:
\begin{eqnarray}
\lambda_1 &=& \beta(1 - \xi) + \frac{\xi}{2} \geq 0, \\
\lambda_2 &=& \gamma(1 - \xi) + \frac{\xi}{2} \geq 0,
\end{eqnarray}
and for the remaining eigenvalues, we have:

\begin{eqnarray}
\lambda_3 + \lambda_4 &=& (\alpha + \delta)(1 - \xi) \geq 0, \\
\lambda_3 \lambda_4 &=& \alpha \delta (1 - \xi)^2 - \frac{\xi^2}{4}.
\end{eqnarray}

From these conditions, we infer that all eigenvalues are non-negative when:
\begin{equation}
\lambda_3 \lambda_4 = \alpha \delta (1 - \xi)^2 - \frac{\xi^2}{4} \geq 0.
\end{equation}
This is equivalent to  
\begin{equation}
r^2 (1 - \xi)^2 - \frac{\xi^2}{4} \geq 0, \quad r \equiv \sqrt{\alpha \delta}.
\end{equation}
We note that $0 \leq r \leq \frac{1}{2}$. Solving the above inequality, we find that all eigenvalues are non-negative, or equivalently, the extended Werner density matrix is separable when  
\begin{equation}
0 \leq \xi \leq \frac{2r}{1 + 2r}
\end{equation}
 A phase diagram of the separability-entanglement phase diagrams is shown in Figure 1.

\begin{figure}[h]
\includegraphics{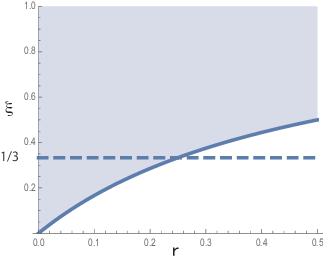}
\caption{Phase diagram showing separability and entanglement regions for the extended Werner density matrix. The white  and grey areas represent the separable and entangled regions, respectively.  The dashed line indicates $\xi = 1/3$.}
\end{figure}

With this generalization, we observe that the maximum value of $\xi$ for which the density matrix remains separable increases from $1/3$ to:
\begin{equation}
\xi_c = \frac{1}{2},
\end{equation}
which is attained when $r = 1/2$, or equivalently, when $\alpha = \delta = 1/2$ and $\beta = \gamma = 0$.

\section{Discussion}

We have demonstrated that a quantum state can still exhibit a separable phase even when described by the extended Werner matrix with a dependent classical component. Furthermore, within a specific parameter range, namely $\frac{1}{4} \leq r \leq \frac{1}{2}$, the separable region - above the dashed line in Figure 1(B) - expands compared to the case with an independent classical component. In other words, introducing classical correlation increases the separability of the state.

Additionally, in the range $0 \leq r \leq \frac{1}{4}$, the entanglement phase is enhanced. However, a similar entanglement effect has been observed when introducing bias while maintaining inependence in the classical component (see Appendix for details).

This leads to a rather peculiar situation: maintaining independence with bias in the classical component enhances entanglement, whereas removing classical independence increases the overall separability of the mixed state. Overall, our results suggest that the conventional view - that the degree of mixing between classical independent and quantum entangled components directly determines the separability - entanglement nature of overall quantum state - does not always hold. In the quantum world, it is possible to extract unentangled strings from a system composed of two entangled groups of strings.


Experimental verification of these findings remains an open challenge.

\section*{Acknowledgments}

The author would like to thank Prof. Emeritus Philip Pearle of Hamilton College and Prof. Yoko Miyamoto of the University of Electro-Communications, Japan, for their careful reading and valuable comments on this work. This work was supported by funding from Ohagi Hospital, Hashimoto, Wakayama, Japan, and by the JSPS Topic-Setting Program to Advance Cutting-Edge Humanities and Social Sciences Research (Grant Number JPJS00122674991), as well as JSPS KAKENHI (Grant Number 19H01201).


\section{Appendix}
We consider a special case of the generalized Werner density matrix in which independence is preserved while introducing bias in the classical component. This case arises when  
\begin{equation}
\alpha = p_A p_B, \quad \beta = p_A (1 - p_B), \quad \gamma = (1 - p_A) p_B, \quad \delta = (1 - p_A) (1 - p_B),
\end{equation}
where  

- A takes state \( a_1 \) with probability \( p_A \) and state \( a_2 \) with probability \( 1 - p_A \),  

- B takes state \( b_1 \) with probability \( p_B \) and state \( b_2 \) with probability \( 1 - p_B \).  
\vspace{1em}

In this case of bias introduction the Werner matrix is given by

\begin{equation}
\hat{\rho}^{W}_{AB} = (1 - \xi) 
\begin{bmatrix}
p_A & 0 \\
0 & 1 - p_A
\end{bmatrix}_A 
\otimes
\begin{bmatrix}
p_B & 0 \\
0 & 1 - p_B
\end{bmatrix}_B
+ \xi \ket{\Psi^{-}} \bra{\Psi^{-}},
\end{equation}
or, in matrix form:
\begin{equation}
\hat{\rho}^{W}_{AB} = 
\begin{bmatrix}
{p_A}{p_B}(1-\xi) & 0 & 0 & 0 \\
0 & {p_A}{(1 - p_B)}(1 - \xi) + \frac{\xi}{2} & -\frac{\xi}{2} & 0 \\
0 & -\frac{\xi}{2} & {(1 - p_A)}{p_B}(1 - \xi) + \frac{\xi}{2} & 0 \\
0 & 0 & 0 & {(1 - p_A)}{(1 - p_B)}(1 - \xi)
\end{bmatrix}.
\label{eWerner}
\end{equation}
We note that we recover the original Werner density matrix when
\begin{equation}
p_A = p_B = \frac{1}{2}.
\label{recover}
\end{equation}

We now apply  the Peres-Horodecki criterion to (\ref{eWerner}) leading to the following condition for separability: 
\begin{equation}
0 \leq \xi \leq \xi_c = \frac{q}{1+q}, \quad q \equiv 2 \sqrt{(1 - p_A)(1 - p_B) p_A p_B}.
\label{biasseparability1}
\end{equation}
We note that $0 \leq q \leq \frac{1}{2}$.

If the bias is identical for A and B, i.e., $p_A = p_B = p$, the separability condition in equation (\ref{biasseparability1}) simplifies to:
\begin{equation}
q = 2 p(1 - p).
\label{biasseparability2}
\end{equation}
The unbiased case $p_A = p_B = \frac{1}{2}$ gives $q = \frac{1}{2}$, recovering the known separability condition for the Werner density matrix:
\begin{equation}
0 \leq \xi \leq \frac{1}{3}.
\end{equation}

We plot the above conditions in Fig. 2, which clearly shows that the region of entanglement is extended beyond $\xi = \frac{1}{3}$.

\begin{figure}[h]
\includegraphics{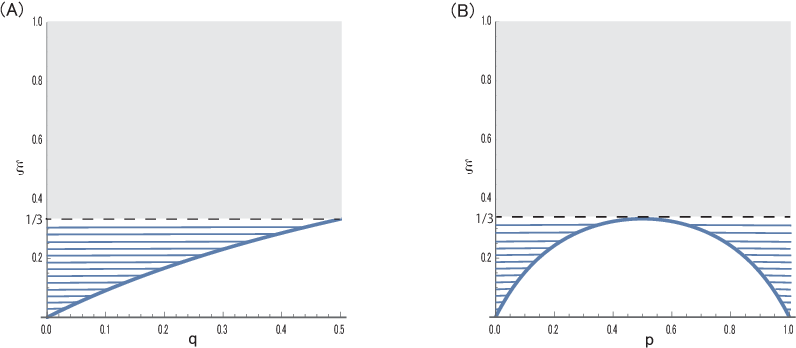}
\caption{Diagrams showing bias-induced entanglement enhancement. The grey area represents the range of entanglement with no bias, while the shaded area shows the enhanced region. (A) General case from equation (\ref{biasseparability1}), (B) The case when the bias is the same, $p_A = p_B = p$, as in equation (\ref{biasseparability2}).}
\end{figure}

We also computed the concurrence of the biased Werner state for different parameter sets. Some results are shown in Fig. 3. From these, we infer that the concurrence of the biased Werner state is given by:

\begin{equation}
C_{\text{bias}}(\xi) = 
\begin{cases}
1 - \frac{1 - \xi}{1 - \xi_c} & \xi_c < \xi \leq 1 \\
0 & 0 \leq \xi \leq \xi_c
\end{cases}.
\label{concurrence}
\end{equation}

We also note that for $\xi > \xi_c$, the concurrence of the biased Werner density matrix $C_{\text{bias}}$ is greater than that of the unbiased Werner density matrix $C_W$:

\begin{equation}
C_W(\xi) < C_{\text{bias}}(\xi), \quad \xi_c < \xi \leq 1.
\label{concurrence2}
\end{equation}

\begin{figure}
\includegraphics[height=4.35cm]{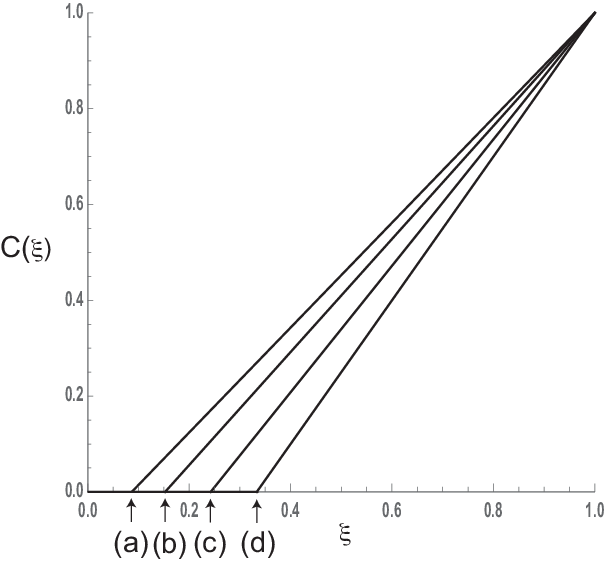}
\caption{Graphs showing bias-induced entanglement enhancement as measured by concurrence, $C(\xi)$. The values of $(q, \xi)$ are (a) $(19/200, 19/219)$, (b) $(9/50, 9/59)$, (c) $(8/25, 8/33)$, and the unbiased case (d) $(1/2, 1/3)$.}
\end{figure}

The above results show that introducing bias while keeping independence into the classical component of the Werner state leads to enhanced entanglement.

\clearpage


\end{document}